\documentclass{article}
\usepackage{spconf,amsmath,graphicx,subcaption,booktabs, multirow}
\usepackage[
singlelinecheck=false % <-- important
]{caption}
\usepackage{import}
\usepackage{gensymb}
\usepackage{cite}
\usepackage{booktabs} 
\usepackage{tabularx}
\usepackage{enumitem}
\usepackage{textcomp}
\usepackage{cite}
\usepackage[english]{babel}
\usepackage{mathtools}

\usepackage{ifpdf}
\usepackage{algorithmic}
\usepackage{multirow}
\usepackage{xcolor}
\usepackage{array}
\usepackage{url}
\usepackage[colorlinks=true, urlcolor=blue, linkcolor=red]{hyperref}
\usepackage{comment}
\usepackage{xurl}

\title{Intelligent Multi-View Test Time Augmentation}
\name{Efe Ozturk, Mohit Prabhushankar, Ghassan AlRegib}
\address { OLIVES at the Center for Signal and Information Processing CSIP,\\ 
School of Electrical and Computer Engineering, Georgia Institute of Technology, Atlanta, GA, USA \\
\{eozturk7, mohit.p, alregib\}@gatech.edu             }

\begin{document}
%\ninept
%

\twocolumn[{%

{ \large
\begin{itemize}[leftmargin=2.5cm, align=parleft, labelsep=2cm, itemsep=4ex,]

\item[\textbf{Citation}]{E. Ozturk, M. Prabhushankar, and G. AlRegib, "Intelligent Multi-View Test Time Augmentation," in \textit{2024 IEEE International Conference on Image Processing (ICIP), Abu Dhabi, United Arab Emirates (UAE), 2024.}}

\item[\textbf{Review}]{Date of Acceptance: June 6th 2024}

\item[\textbf{Codes}]{\url{https://github.com/olivesgatech/Intelligent-Multi-View-TTA}}

\item[\textbf{Bib}]  {@inproceedings\{ozturk2024intmultiviewtta,\\
    title=\{Intelligent Multi-View Test Time Augmentation\},\\
    author=\{Ozturk, Efe and Prabhushankar, Mohit and AlRegib Ghassan\},\\
    booktitle=\{IEEE International Conference on Image Processing (ICIP)\},\\
    year=\{2024\}\}}

\item[\textbf{Copyright}]{\textcopyright 2024 IEEE. Personal use of this material is permitted. Permission from IEEE must be obtained for all other uses, in any current or future media, including reprinting/republishing this material for advertising or promotional purposes, creating new collective works, for resale or redistribution to servers or lists, or reuse of any copyrighted component of this work in other works.}

\item[\textbf{Contact}]{

\{eozturk7, mohit.p, alregib\}@gatech.edu \\\url{https://ghassanalregib.info/}\\}
\end{itemize}

}}]

\maketitle
%
%\section{Abstract}
\begin{abstract}
In this study, we introduce an intelligent Test Time Augmentation (TTA) algorithm designed to enhance the robustness and accuracy of image classification models against viewpoint variations. Unlike traditional TTA methods that indiscriminately apply augmentations, our approach intelligently selects optimal augmentations based on predictive uncertainty metrics. This selection is achieved via a two-stage process: the first stage identifies the optimal augmentation for each class by evaluating uncertainty levels, while the second stage implements an uncertainty threshold to determine when applying TTA would be advantageous. This methodological advancement ensures that augmentations contribute to classification more effectively than a uniform application across the dataset. Experimental validation across several datasets and neural network architectures validates our approach, yielding an average accuracy improvement of 1.73\% over methods that use single-view images. This research underscores the potential of adaptive, uncertainty-aware TTA in improving the robustness of image classification in the presence of viewpoint variations, paving the way for further exploration into intelligent augmentation strategies. The code is available at: \href{https://github.com/olivesgatech/Intelligent-Multi-View-TTA}{https://github.com/olivesgatech/Intelligent-Multi-View-TTA}
\end{abstract}


\vspace{-2mm}
\section{Introduction}
\vspace{-2mm}
Deep Neural Networks (DNNs) have demonstrated unprecedented success in image classification, surpassing human benchmarks on pristine data\cite{he2016deep}. Despite these advancements, DNNs remain sensitive to variations in input data, such as noise \cite{temel2017cure} and shifts in the viewpoint of objects \cite{temel2018cure}. These variations can reduce the efficacy of DNNs in real-world applications where ideal conditions are rarely met~\cite{prabhushankar2022introspective}.

\begin{figure}
    \centering   \includegraphics[height=2.5cm]{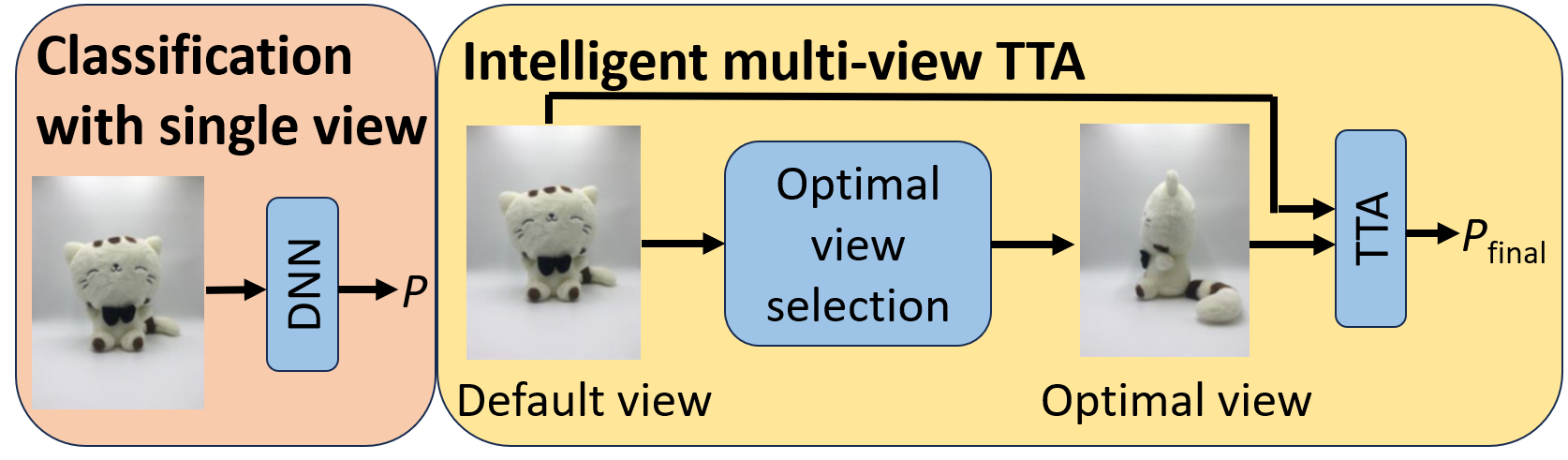}
    \caption{Comparison of Intelligent Multi-View TTA with the conventional single-view method. This illustrates how the intelligent approach dynamically selects augmentation views to refine predictions (P), in contrast to the conventional method's reliance on a single, static view.}
    \label{fig:int_tta}
\end{figure}

Data augmentation is a widely recognized strategy to enhance the robustness and generalization capability of DNNs by artificially increasing the diversity of training data \cite{shorten2019survey}. Traditional data augmentation techniques, applied during the training phase, include transformations like rotations, scaling, and color adjustments to simulate variability in the input data. Test Time Augmentation (TTA) extends this concept to the inference phase, applying augmentations to test data to improve model predictions \cite{kimura2021understanding}. TTA is based on the idea that evaluating multiple, varied versions of the same input can provide a more comprehensive representation, leading to more accurate and robust predictions.

However, conventional data augmentation methods often apply augmentations uniformly without considering each class's specific characteristics or challenges. This approach can lead to suboptimal or detrimental effects on classification performance, particularly when irrelevant or misleading augmentations are applied. Therefore, adopting a class-based data augmentation strategy is essential for enhancing classification performance by ensuring that augmentations are relevant and tailored to the specific needs of each class \cite{yoo2023class}.

\begin{figure*}
    \centering
    \includegraphics[width=0.9\linewidth]{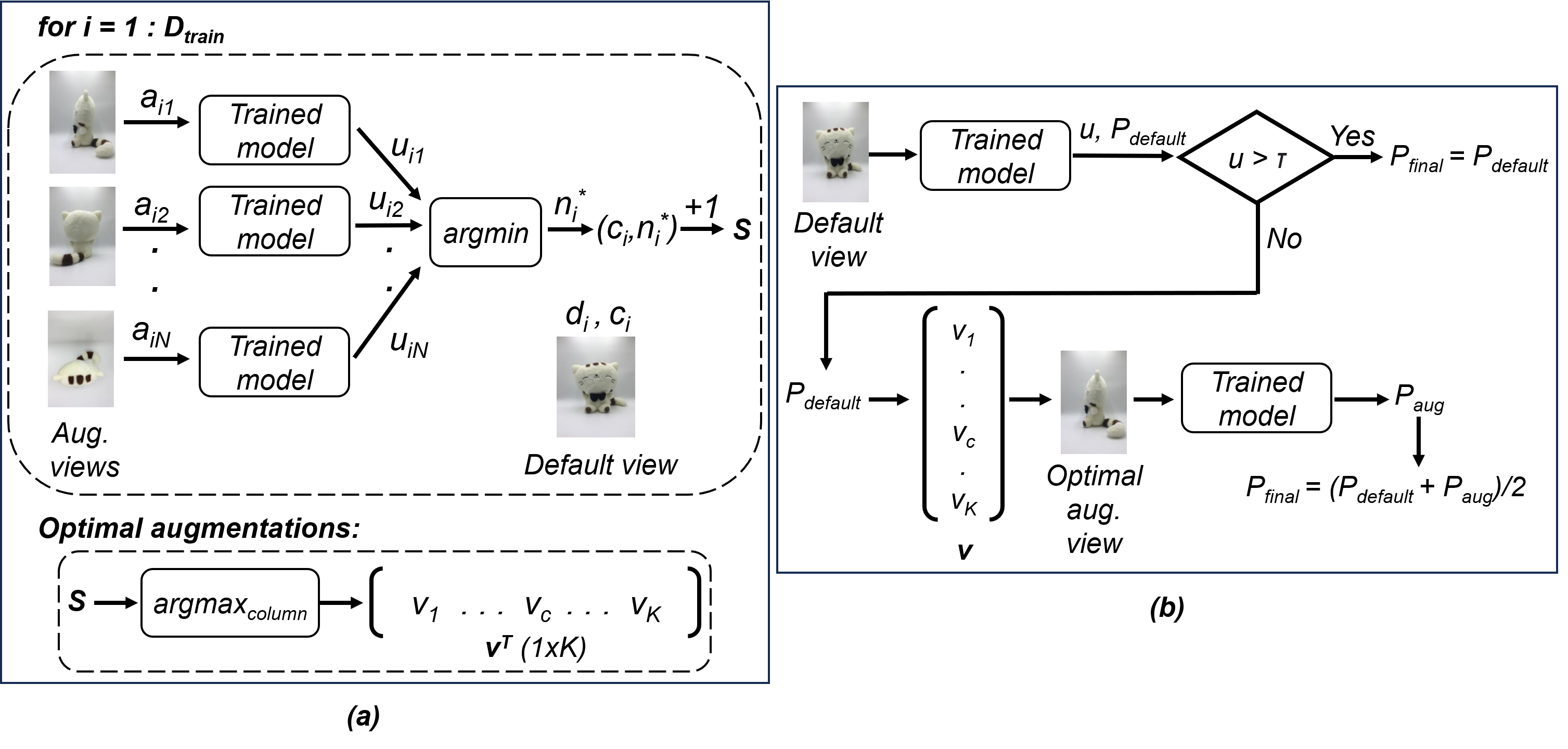}
    \caption{The proposed algorithm. (a) Stage-1: optimal augmentation view selection. (b) Stage-2: uncertainty assessment.}
    \label{fig:algorithm}
\end{figure*}

Addressing this gap, our study introduces a TTA algorithm (Fig.~\ref{fig:int_tta}) that enhances image classification robustness specifically against variations in object viewpoints, a common and challenging source of variation in real-world scenarios. Our method diverges from standard practices by employing predictive uncertainty to select the most appropriate viewpoint augmentation for each class intelligently. Through a two-stage process, we first identify the optimal augmentation view for each class based on uncertainty metrics. We then apply an uncertainty assessment mechanism to determine when the use of TTA is likely beneficial, ensuring that augmentations are applied in a targeted and efficient manner.

By focusing on variations in viewpoints and leveraging predictive uncertainty, our algorithm represents a significant advancement in applying TTA for image classification. Our extensive experiments, conducted across multiple datasets and neural network architectures, demonstrate an average accuracy improvement of 1.73\% over DNNs that use a single view of objects. Furthermore, our method achieves an average accuracy improvement of 10.45\% compared to applying random augmentations uniformly across the entire dataset. This highlights the critical importance of selecting optimal augmentations and employing them in a targeted manner. These findings underscore the potential of our approach to substantially improve the robustness and accuracy of DNNs in handling real-world image classification tasks.

\vspace{-3mm} 
\section{Literature Review}

In our pursuit of addressing the robustness challenges in image classification, our choice to concentrate on multiple views as the augmentation type is grounded in a growing body of research highlighting the advantages of leveraging diverse perspectives. Several studies have demonstrated that utilizing different views of an object outperforms methods that rely on a single view \cite{seeland2021multi,zhang2019multi}. By incorporating multiple views of objects, DNNs can harness a variety of perspectives, enabling them to develop a more comprehensive understanding of the object's features and enhancing their ability to generalize across varying orientations. This insight drives our focus on utilizing multiple views of objects as a promising augmentation strategy within the TTA framework.

One of the central challenges in multi-view analysis, which includes our task of selecting the optimal augmentation view, lies in the relationship between object classes and the most informative views. While the merits of employing multiple views of objects are well-documented \cite{li_2019_a}, determining which view best represents a specific class remains a complex task. The challenge arises from the fact that the optimal view for one class may not be the same as the optimal view for another. This inherent class-specificity complicates the augmentation selection process and necessitates a tailored approach to address this challenge.

Our study concentrates on three distinct areas spanning diverse domains: object recognition, face recognition, and medical imaging. In the domain of object recognition, the CURE-OR dataset \cite{temel2018cure} comprises a diverse collection of 100 daily objects. Each object is represented by five distinct views: front, left, back, right, and top views. This dataset focuses on object recognition, presenting a variety of daily objects captured under varying conditions, including noise, background variations, and different camera angles. It effectively addresses challenges related to robustness in object recognition. In the context of our study, this dataset provides information from multiple views, allowing us to gain a more comprehensive understanding of objects. Furthermore, determining which views are most informative is essential for optimizing image classification accuracy. Therefore, the CURE-OR dataset offers a valuable context for evaluating the effectiveness of our proposed algorithm in object classification scenarios.

\begin{figure}
    \centering
    \includegraphics[width=7cm]{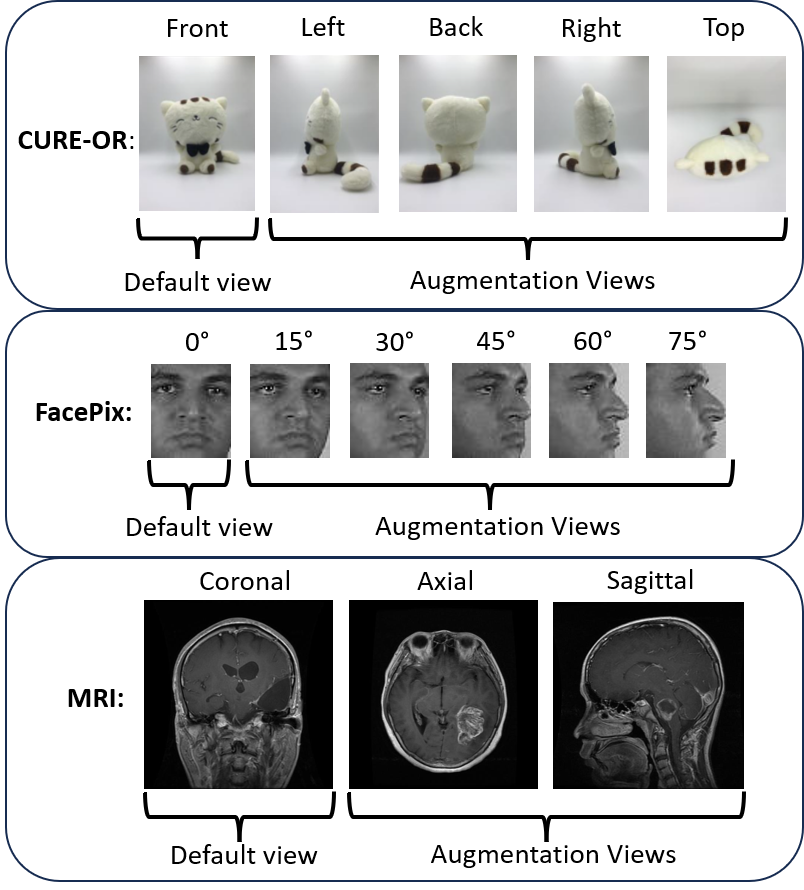}
    \caption{Example images from the datasets CURE-OR, FacePix and Brain Tumor MRI. Default view and augmentation views are indicated for each of the datasets.}
    \label{fig:datasets}
\end{figure}

Face recognition finds widespread applications in various domains, including commercial products and security systems. The angle of the subject's face can significantly impact the performance of these applications, highlighting the need for increased robustness \cite{zhang2009face}. In this context, the FacePix dataset \cite{little2005methodology} offers faces captured from multiple angles, with a specific emphasis on facial features and devoid of additional contextual information. This dataset's deliberate focus on diverse facial angles aligns seamlessly with our research goals. The FacePix dataset comprises images of faces from 30 different individuals, with each individual's face captured across 181 distinct head poses. Specifically, the captures cover a range of yaw angles, with 1\degree increments spanning from -90\degree to 90\degree. By identifying the optimal view for face recognition within this extensive dataset, we aim to elevate the accuracy of facial feature recognition, effectively tackling the unique challenges arising from varying angles and perspectives.

In the domain of medical imaging, MRI (Magnetic Resonance Imaging) is widely utilized, playing a pivotal role in healthcare diagnostics \cite{katti2011magnetic}. Producing each MRI image, often called a slice, requires substantial time and resources. Therefore, identifying the optimal view for classification can significantly streamline the efforts involved in acquiring and processing MRI images. For this purpose, we use the Brain Tumor MRI dataset \cite{Nickparvar_2021}. This dataset provides a diverse array of medical imaging perspectives, capturing brain tumors from different slices and angles in a computed 360-degree fashion. The Brain Tumor MRI dataset consists of three distinct tumor classes: Glioma, Meningoma, and Pituitary, with each image acquired from three primary views: coronal, axial, and sagittal. This dataset enriches our investigation by focusing on medical imaging scenarios, and the application of our approach to this dataset is aimed at enhancing the efficiency of medical image classification tasks.

Our motivation for addressing the robustness in image classification through TTA draws inspiration from several works emphasizing the necessity of an intelligent and optimal augmentation strategy. Kim et al. \cite{kim2020learning} propose an instance-aware TTA algorithm that dynamically selects suitable transformations based on predicted losses, showcasing efficiency and effectiveness in ensembling. Molchanov et al. \cite{lyzhov2020greedy} introduce the Greedy Policy Search algorithm, providing valuable insights into augmentation policy development. Shanmugam et al. \cite{shanmugam2021better} identify suboptimal methods in traditional TTA and propose a learning-based approach with optimal weights per augmentation. Mocerino et al. \cite{mocerino2021adaptta} present AdapTTA, influencing our concept of uncertainty assessment through its adaptive mechanism.

Furthermore, our approach incorporates the concept of uncertainty quantification, inspired by active learning techniques. Uncertainty quantification deals with assigning probabilities regarding decisions made under some unknown states of the system. Quantifying uncertainty is an essential step in multiple applications including out-of-distribution detection~\cite{lee2022gradient}, visual explanation analysis~\cite{prabhushankar2024voice} and adversarial image detection~\cite{lee2023probing}. In active learning, uncertainty is a commonly used criterion for selecting data during the training phase \cite{lewis1994heterogeneous}. Active learning algorithms often prioritize selecting the most uncertain data points, leveraging them in the training set to improve model performance. Benkert et.al~\cite{benkert2023gaussian} count the number of times the class predictions of data switches during training to estimate uncertainty. In the context of TTA, we extend the use of existing predictive uncertainty metrics in two crucial aspects: firstly, in the selection of optimal views during the augmentation process, and secondly, in the inference phase to assess the necessity of TTA. Leveraging established uncertainty metrics allows us to systematically identify and select optimal views and assess the need for TTA, enhancing both the robustness and accuracy of our image classification model.

\vspace{-3mm}
\section{Methodology}
\vspace{-3mm}
The Intelligent Multi-View TTA algorithm consists of two stages, each of which is shown in Fig.~\ref{fig:algorithm} and detailed below.

\vspace{-3mm} 
\subsection{Stage-1: Optimal Augmentation View Selection}

Let $\mathcal{D}$ be a dataset of images with $K$ classes. For each image $i \in \mathcal{D}$, there exists a default view $d_i$ and a set of $N$ augmentation views $A_i = \{a_{i1}, a_{i2}, \ldots, a_{iN}\}$, where $N$ is the total number of augmentation views available for each image. We split $\mathcal{D}$ into training and test sets, $\mathcal{D}_{train}$ and $\mathcal{D}_{test}$, respectively. We train a DNN using default views $d_i$ from $\mathcal{D}_{train}$, and label it as the trained model. For each $d_i$ in $\mathcal{D}_{train}$, we determine the optimal augmentation from $A_i$ by calculating the predictive uncertainty $u_{ij}$ for each augmentation view $a_{ij}$, where $j$ represents the index within the set of possible augmentations $N$. The predictive uncertainty $u_{ij}$ quantifies the model's confidence in its prediction for the view $a_{ij}$. The optimal augmentation for $d_i$ is determined as:
\begin{equation}
n_i^* = \underset{j \in {1, \ldots, N}}{\text{argmin }} u_{ij}.
\label{eq:eq1}
\end{equation}

To systematically record the optimal augmentations, we construct the matrix $\mathbf{S}$ of dimensions $K \times N$, where each entry $\mathbf{S}_{c,n}$ represents the frequency with which an augmentation view $n$ is selected as optimal for images belonging to class $c$. Initially, all entries of $\mathbf{S}$ are set to zero. Upon identifying $n_i^*$ for an image $d_i$ of class $c_i$, we increment $\mathbf{S}_{c_i,n_i^*}$ by 1. After processing all images in $\mathcal{D}_{train}$, the algorithm calculates the optimal augmentation for each class, encapsulated within the vector $\mathbf{v}$ of dimensions $K \times 1$. This is achieved by:
\begin{equation}
\mathbf{v}_{c} = \underset{n}{\text{argmax }} \mathbf{S}_{cn} \quad \forall c \in \{1, \ldots, K\}.
\label{eq:eq2}
\end{equation}
In Equation~\ref{eq:eq2}, $\mathbf{v}_{c}$ denotes the index of the augmentation, which is selected as optimal for class $c$.

\begin{figure}
    \centering
    \includegraphics[width=7cm]{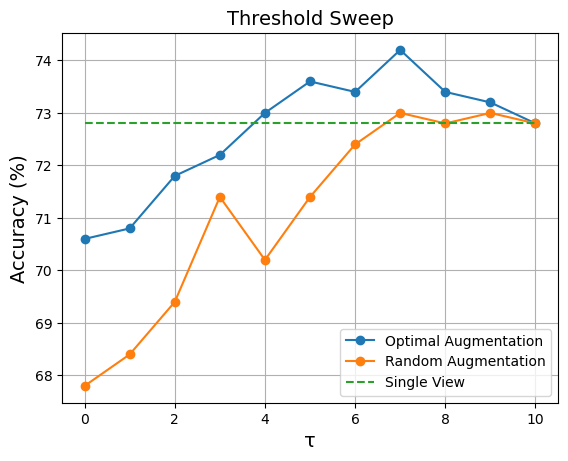}\vspace{-2mm}
    \caption{Threshold sweep of entropy for VGG-16 trained on CURE-OR.}
    \label{fig:sweep}
\end{figure}

\vspace{-0.3cm}

\subsection{Stage-2: Uncertainty Assessment}
In the second stage of our algorithm, we assess the necessity of applying TTA through an uncertainty evaluation process for the default view images in $\mathcal{D}_{test}$. For each image, the trained model generates a softmax prediction vector, $P_{default}$, of dimensions $K \times 1$, where $K$ is the number of classes. Alongside this, the corresponding predictive uncertainty $u$ is calculated. If $u$ surpasses the predefined threshold $\tau$, indicating a lack of confidence in the model’s initial prediction, the algorithm references the vector $\mathbf{v}$ to guide the augmentation selection process. Specifically, the algorithm identifies the class $c$ that has the highest probability in $P_{default}$ and consults $\mathbf{v}_c$ to determine the index of the optimal augmentation view. The view with the index $\mathbf{v}_c$ is then used as the optimal augmentation. A new prediction vector, $P_{aug}$, is obtained from the optimal augmentation, and the final prediction, $P_{final}$, is calculated by averaging $P_{default}$ and $P_{aug}$. 
Conversely, if the uncertainty is below $\tau$, indicating high confidence in the initial prediction, $P_{default}$ is directly utilized as $P_{final}$, without further augmentation.

\begin{table}
\caption{Accuracy of single view, random augmentation, and intelligent TTA across datasets and models.}
\centering
\begin{tabular}{ccccc}
\toprule
\multirow{2}{*}{Dataset} & \multirow{2}{*}{Model} & \multicolumn{3}{c}{Accuracy (\%)} \\
\cmidrule{3-5}
& & \begin{tabular}{@{}c@{}}Single\\ View\end{tabular} & \begin{tabular}{@{}c@{}}Random\\ Aug.\end{tabular} & \begin{tabular}{@{}c@{}}Int.\\ TTA\end{tabular} \\
\midrule
\multirow{3}{*}{CureOr} & ResNet50 & 74.20 & 72.89 & 75.90 \\
                        & VGG16 & 72.80 & 67.93 & 73.37 \\
                        & VITB16 & 76.60 & 73.56 & 76.97 \\
\midrule
\multirow{3}{*}{FacePix} & ResNet50 & 78.79 & 58.08 & 83.84 \\
                         & VGG16 & 80.30 & 78.28 & 82.32 \\
                         & VITB & 81.82 & 50.75 & 83.33 \\
\midrule
\multirow{3}{*}{MRI} & ResNet50 & 82.60 & 85.74 & 86.71 \\
                     & VGG16 & 84.06 & 72.70 & 84.06 \\
                     & VITB16 & 81.16 & 73.91 & 81.40 \\
\bottomrule
\end{tabular}
\label{table:small}
\end{table}

\vspace{-2mm}
\section{Experiments and Results}
\vspace{-2mm}
In this section, we present the results of experiments on the proposed algorithm, demonstrating its effectiveness in image classification through comparisons with various metrics.

\begin{table*}[ht]
\caption{Minimum vs. optimized threshold results across uncertainty metrics, datasets, and models.}
  \centering
\begin{tabular}{ccccccccccc}
\toprule
\multirow{2}{*}{Dataset} & \multirow{2}{*}{Model} &  \multicolumn{6}{c}{Accuracy (\%)} \\
\cmidrule(lr){3-8}
& &  Entropy  &   NLL &  Brier &  ODIN &   MCD &  GradNorm \\
 & & & & & & & & \\
\midrule
\multirow{6}{*}{} {CURE-OR}&ResNet50     &	\textbf{76.00}/\textbf{77.40}	&68.60/74.20	&\textbf{77.00}/\textbf{78.60}&	70.20/74.20	&\textbf{75.00}/\textbf{75.80}	&\textbf{75.00}/\textbf{75.20} \\
&VGG-16    &70.60/\textbf{74.20}	&	61.80/72.80	&70.20/\textbf{73.40}&	62.60/72.80&	72.00/\textbf{74.40}		&68.20/72.60 \\
&VIT\textunderscore B\textunderscore16     &	75.40/\textbf{77.40}	&	71.40/76.60	&75.00/\textbf{77.20}	&71.80/76.60	&75.20/\textbf{77.40}	&73.80/76.60 \\

\midrule
\multirow{6}{*}{} {FacePix}&ResNet50     	&\textbf{87.88}/\textbf{89.40}	&	39.39/78.79	&\textbf{87.88}/\textbf{87.88}&	66.67/78.79	&\textbf{87.88}/\textbf{89.40}	&40.91/78.79 \\
&VGG-16    	&74.24/\textbf{83.33}		&\textbf{81.82}/\textbf{81.82}	&75.76/\textbf{81.82}	&80.30/\textbf{81.82}	&74.24/\textbf{83.33}	&80.30/\textbf{81.82} \\
&VIT\textunderscore B\textunderscore16     	&72.73/\textbf{83.33}		&48.48/\textbf{83.33}	&72.73/\textbf{84.85}	&48.48/\textbf{83.33}	&72.73/\textbf{83.33}		&48.48/\textbf{81.82} \\
\midrule
\multirow{6}{*}{} {MRI}&ResNet50     	&81.16/\textbf{85.51}		&\textbf{89.86}/\textbf{89.86}	&81.16/\textbf{85.51}	&\textbf{86.96}/\textbf{86.96}	&81.16/\textbf{85.51}	&\textbf{86.96}/\textbf{86.96} \\
&VGG-16    	&73.91/82.60		&82.60/\textbf{85.51}	&73.91/84.06	&65.22/84.06	&63.77/84.06		&71.01/84.06 \\
&VIT\textunderscore B\textunderscore16     	&72.46/81.16		&75.36/\textbf{84.06}	&75.36/79.71	&72.46/81.16	&72.46/81.16	&78.26/81.16 \\
\bottomrule
\end{tabular}
\label{table:big}
\end{table*}

For each dataset, we designate a specific view as the default, with the remaining views utilized for augmentation. In the CURE-OR dataset, we choose the front view as the default and employ the left, back, top, and right views as augmentations. For FacePix, we utilize images with yaw angles spaced 15\degree apart, starting from 0\degree and culminating at 75\degree. Here, the 0\degree view is designated as the default, with the subsequent angles serving as augmentations. In the MRI dataset, the coronal view is selected as the default, while the axial and sagittal views are designated as augmentations. Fig.~\ref{fig:datasets} shows a sample image from each dataset, indicating the default and augmentation views. We split each dataset into training and test sets, with an 80\% and 20\% split ratio, respectively. As DNNs, we utilize fine-tuned versions of ResNet50, VGG-16, and Vision Transformer (VIT\textunderscore B\textunderscore16). These models incorporate modifications in their fully connected layers, including adding three sequential fully connected layers with ReLU activation and dropout, followed by a final layer utilizing softmax activation for classification. We initialize these fine-tuned models with weights pre-trained on the ImageNet dataset and train them using the default view images from the training set.

Using these trained models, we derive two reference classification accuracy metrics. The first metric, 'single view accuracy,' is obtained by evaluating the trained models with the default view images from the test set. This metric serves as a baseline, indicating the performance of the trained models on single views. The second accuracy metric, labeled 'random augmentation accuracy,' is derived by applying TTA to all default view images in the test set using a randomly selected augmentation view. We then calculate the accuracy based on the prediction results obtained with TTA. Random augmentation accuracy is a reference point to evaluate TTA performance when random augmentations are uniformly applied across the dataset, in contrast to employing optimal augmentations selected by our algorithm specifically for images with high uncertainty.

To calculate the predictive uncertainty values used in both Stage-1 and Stage-2 of the algorithm, we employ various uncertainty metrics from the literature, which are entropy \cite{shannon1948mathematical}, negative-log likelihood (NLL) \cite{zhu2018negative}, Brier score \cite{brier1950verification}, ODIN  score \cite{liang2017enhancing}, Monte Carlo Dropout (MCD) \cite{gal2016dropout}, and GradNorm \cite{chen2018gradnorm}. Using multiple uncertainty metrics allows us to calculate an average accuracy result across all metrics and yield insights into each metric's performance.

We start the evaluation of our algorithm's performance by comparing outcomes using optimal augmentation against random augmentation. Lacking a predefined $\tau$ as the uncertainty threshold for Stage-2, we assess the performance across various $\tau$ values. To this end, we sweep the $\tau$ value for each uncertainty metric. This sweep covers the range from the minimum to the maximum uncertainty values observed for the default view images of the test set, selecting 11 equidistant $\tau$ values within this range. For each $\tau$, predictions are generated for all images. If the uncertainty of a default view image surpasses $\tau$, TTA is applied; otherwise, the prediction from the default view serves as the final decision. Classification accuracy is then calculated for each $\tau$ value. 
In Fig.~\ref{fig:sweep}, we present the $\tau$ sweep graph for the VGG-16 model, which was trained with the CURE-OR dataset with entropy as the uncertainty metric. The horizontal axis denotes the series of equidistant $\tau$ values, ranging from 0 to 10, while the vertical axis displays the corresponding accuracy percentages. Due to the variability in the range of values across different uncertainty metrics, we label the $\tau$ values as $0^{th}$, $1^{st}$, etc., without specifying exact numbers. This graph compares the effects of applying TTA with optimal and random augmentations. A constant dashed line is also included to represent the baseline accuracy of the single view for comparison. It is shown that the optimal and random view accuracies start at the lowest level, where TTA is applied for all the images in the test set. By the $4^{th}$ $\tau$ value, the accuracy of the optimal augmentation view surpasses that of the single view. As the $\tau$ increases, accuracies rise to a peak, suggesting the existence of a threshold that maximizes accuracy before eventually aligning with the single view accuracy when TTA is no longer applied at the highest $\tau$. The consistently higher performance of optimal augmentation across all $\tau$ values validates that Stage-1 effectively selects augmentations, leading to better accuracy than random augmentations. Furthermore, the graph confirms that the algorithm achieves the maximum accuracy of 74.2\% at $7^{th}$ $\tau$ value for this specific dataset and model, enhancing the single view accuracy by 1.4\%.

Table \ref{table:small} presents the reference accuracy metrics (Single View, Random Aug.) alongside the accuracy results of our algorithm (Int.\ TTA) across the datasets and models. The Int.\ TTA accuracy is determined by averaging the highest accuracy values achieved by our algorithm across all employed uncertainty metrics.

When comparing the single view and random augmentation accuracies, it becomes evident that random augmentations do not enhance performance; in fact, they lead to a degradation in performance for all datasets. Specifically, we note performance declines of 3.07\%, 17.93\%, and 5.16\%  on average for CURE-OR, FacePix, and MRI datasets, respectively. This observation validates our rationale for adopting an optimal view selection strategy, as it shows that random augmentation fails to contribute positively to model accuracy, underscoring the necessity for a more discerning approach.

Furthermore, the comparison between Int.\ TTA and random augmentation accuracies reveal substantial improvements, with average increases of 3.95\%, 20.8\%, and 6.6\% for CURE-OR, FacePix, and MRI datasets, respectively. These figures not only underscore the inefficacy of random augmentation but also highlight the significant advantages of our algorithm's approach. By meticulously selecting the optimal augmentation view for each class, especially in the context of the challenges specific to each dataset, our algorithm ensures a consistent elevation in classification results over the random augmentation approach.

Comparing Int.\ TTA accuracy with single-view accuracy reveals significant enhancements across the datasets, underlining the efficacy of our algorithm against the traditional single-view approach. In the CURE-OR dataset, characterized by its high variability in object orientation, our algorithm markedly improves classification accuracy. Specifically, the ResNet50 model sees an increase from the single view accuracy of 74.20\% to 75.9\% with Int.\ TTA. This specific increase, along with an average enhancement of 0.87\% across the models, emphasizes the algorithm's capability to adeptly navigate the complexities of varying object orientations.

For the FacePix dataset, which presents challenges in recognizing facial features from different angles, our algorithm demonstrates substantial effectiveness. The ResNet50 model experienced a substantial boost in accuracy, from 78.79\% to 83.84\%. This marked increase, coupled with an average accuracy gain of 2.87\% across the models, highlights the strength of our algorithm in significantly improving facial recognition accuracy amidst angular variations.

The MRI dataset, demanding precise classification across different medical imaging views, also benefited from our algorithm's approach. The most considerable impact was observed with the ResNet50 model, where accuracy was enhanced from 82.60\% to 86.71\%. Although this improvement averages at 1.45\% across the models, such an enhancement holds particular importance in medical imaging, where minor accuracy gains can translate into considerable diagnostic advancements. Furthermore, our algorithm's efficiency in enhancing accuracy by using only the default and optimal augmentation views underscores a critical advantage: there is no need to utilize images from all views. This approach addresses the challenge of effectively using time and computing resources during the acquisition of MRI images, making the process more streamlined and resource-efficient.

In Table \ref{table:big}, we present the accuracy outcomes of our algorithm under various uncertainty metrics, showcasing the efficacy of our algorithm. Each cell's first value indicates the accuracy achieved at the minimum $\tau$, applying optimal augmentations across all test images. The second value represents the peak accuracy within the $\tau$ sweep range. Typically, the latter value is higher, demonstrating that incorporating an uncertainty assessment mechanism significantly boosts performance. We highlight instances where the results exceed the baseline single-view accuracy, illustrating the algorithm's precision in enhancing model efficacy. For the CURE-OR dataset with the ResNet50 model, employing the Brier score as the uncertainty metric resulted in the most substantial accuracy gain of 4.4\% over the corresponding single view accuracy. Similarly, in the FacePix dataset, entropy and MCD metrics led to significant increases, with the ResNet50 model showing an improvement of 10.61\%. In the MRI dataset context, NLL offered the most pronounced enhancement, achieving a 7.26\% increase for the ResNet50 model. These findings underscore the algorithm's capacity to not only optimize image classification accuracy but also its adaptability across different datasets and uncertainty metrics, proving the method's broad applicability and effectiveness.

\vspace{-2mm}
\vspace{-2mm}
\section{Conclusion}
\vspace{-2mm}
In this study, we introduced the intelligent multi-view TTA algorithm that enhances the robustness and accuracy of image classification models, particularly in the face of viewpoint variations. By intelligently selecting optimal augmentation views and when to use them based on predictive uncertainty, our approach diverges from traditional TTA methods that apply augmentations uniformly, marking a significant advancement in the application of TTA. Our comprehensive experiments across multiple datasets and using various deep learning architectures demonstrate the superiority of our method. The algorithm consistently improved classification accuracy, outperforming both the models using single-view and those using random augmentations in TTA. The results underscore the efficacy of utilizing predictive uncertainty to guide the selection of augmentations, ensuring that TTA is applied in a manner that is both targeted and beneficial.

%Looking forward, the principles established by our work offer a promising foundation for exploring additional augmentation types and strategies beyond viewpoint variations. Future studies could investigate the integration of our intelligent TTA approach with emerging deep learning models and architectures, further enhancing their performance and generalizability. Additionally, exploring the scalability of our method in larger, more complex datasets and real-world scenarios will be crucial for validating its applicability and effectiveness in diverse applications.

\vspace{-3mm}
\bibliographystyle{IEEEbib}
\bibliography{refs}

\end{document}